# Substrate contact angle governs microgel shape, stiffness and deposition pattern

M. Friederike Schulte[1], Sebastian S. Meyer[1], Timon Kratzenberg[2], Simon Schog[2], Jesco M. Schönfelder[1], Silke Klein-Kormelink[1], Tim Blinzer[1], Matthias Karg[3], Walter Richtering[2], Marcel Rey[1]

[1] Institute of Physical Chemistry, University of Münster, Germany

[2] Institute of Physical Chemistry, RWTH Aachen University, Germany

[3] Physical Chemistry of Functional Polymers, Martin Luther University Halle-Wittenberg, Germany

Corresponding author: Marcel Rey; mrey@uni-muenster.de

## Abstract

Soft microgels are widely used as deformable building blocks for two-dimensional assemblies, yet solid substrates are often treated as passive supports after interfacial deposition. Here, we show that substrate wettability mechanically preconditions soft microgels before drying. Using in-liquid force-volume atomic force microscopy, we find that the same microgels adopt markedly different hydrated shapes and stiffness profiles depending on substrate contact angle: hydrophobic substrates induce spreading, flattening, and internal stiffening, whereas hydrophilic substrates preserve taller, softer microgels with smaller contact areas. These single-microgel states can explain how Langmuir-Blodgett-deposited monolayers respond during drying. On hydrophilic substrates, the observed assemblies are consistent with soft and weakly immobilized microgels rearranging under immersion-capillary forces, producing distinct corona-corona and core-core contact states and an apparent isostructural transition. On hydrophobic substrates, the flattened and stiffened microgels are more strongly immobilized, likely suppressing capillary-driven rearrangements and largely preserving the transferred interfacial assembly structure. These findings establish substrate-controlled microgel mechanics as the missing link between interfacial self-assembly and the final structures observed after transfer and drying.

## Keywords

Microgels, atomic force microscopy, stiffness tomography, substrate contact angle, two-dimensional phase behavior, isostructural phase transition

## TOC

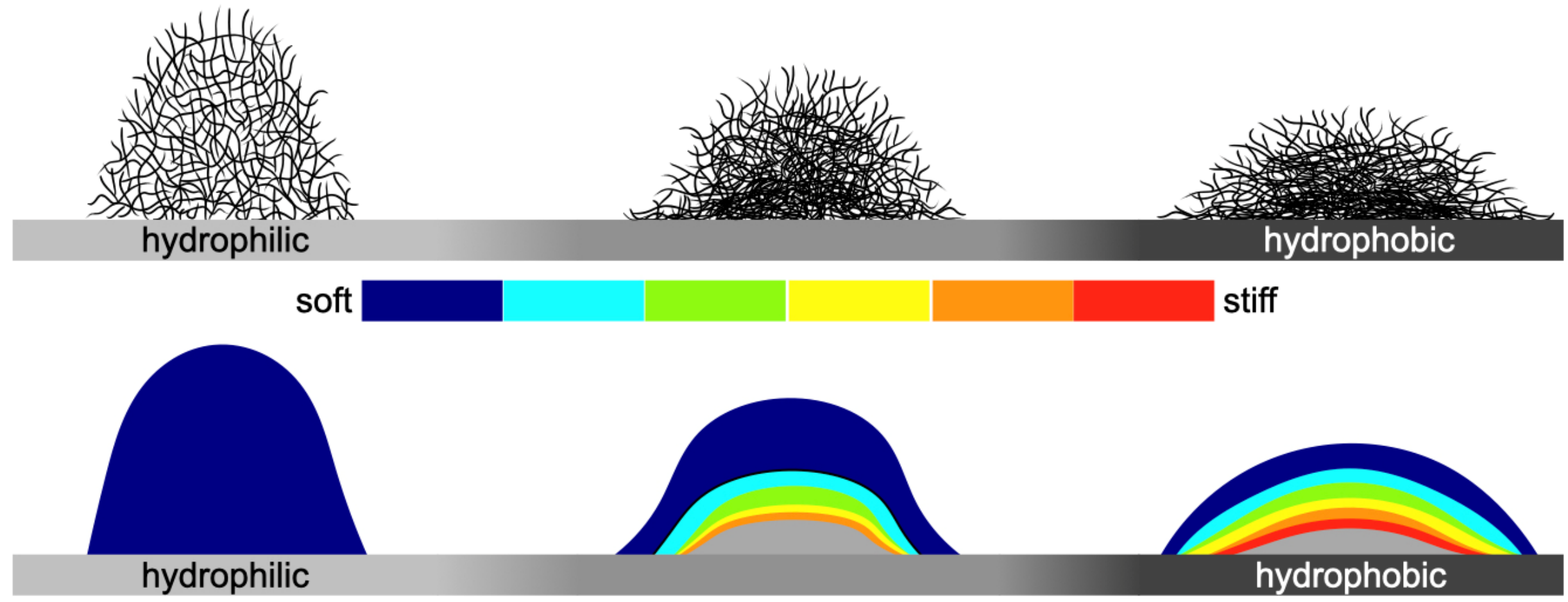

## Introduction

Microgels are soft colloidal particles made of cross-linked polymer networks.[1–3] Despite their non-amphiphilic nature, they exhibit pronounced interfacial activity. [1–4] Upon adsorption to interfaces, they spread and adopt distinct interfacial morphologies,[5–11] while simultaneously modifying the structure and rheology of the liquid interface.[12,13] This combination makes microgels versatile building blocks for adaptive soft materials,[14] including stimuli-responsive emulsions[15–21] and foams,[22–24] membranes[14,25], functional substrates,[25–30] and surface patterning strategies.[31–35] In these systems, interfacial properties can be dynamically tuned by external stimuli such as temperature,[15,17,18,20,25,27,28,30,36] pH,[15–17,37] or ionic strength.[37–39]

At the same time, interfacial adsorption confines microgels and core-shell particles to two dimensions, rendering them attractive model systems for collective phenomena of soft, deformable colloids. Depending on particle architecture, different assembly behaviors have been reported. Core-shell particles with weakly cross-linked or effectively un-cross-linked shells undergo symmetry-changing transitions between hexagonal, chain-like, and square lattices, observed both *in situ* and *ex situ*.[40,41] In contrast, microgels and conventionally cross-linked core-shell particles have predominantly been reported to exhibit an isostructural solid-solid transition upon compression,[32,42–48] characterized by a change from corona-corona to core-core contact within a hexagonally ordered monolayer. This phenomenon has been observed in cryo-SEM[4,13,49,50] at the interface of oil/water emulsions as well as *ex situ* after deposition onto a solid substrate.[32,42–48]

Compelling *in situ* measurements at liquid interfaces, however, show that this isostructural transition is absent under compression and emerges only after transfer to solid substrates.[51,52] Recent work demonstrates that the apparent transition originates from drying-induced rearrangements driven by attractive immersion capillary forces rather than from an intrinsic interfacial phase behavior.[53] The manifestation of these transitions is strongly substrate dependent: they are observed on hydrophilic substrates that permit high lateral mobility during solvent evaporation but are suppressed on hydrophobic substrates, where strong microgel-substrate adhesion quenches rearrangements.[53] Together, these findings support the view that the widely reported isostructural transition does not reflect an intrinsic interfacial phase behavior, but instead arises from transfer- and drying-induced rearrangements during sample preparation on solid substrates.

Taken together, the strong substrate dependence of drying-induced rearrangements implies that the substrate influences not only microgel mobility during solvent evaporation but also the mechanical state of the microgels themselves. Consistent with this interpretation, super-resolution fluorescence microscopy has shown that the substrate contact angle strongly governs microgel conformation at interfaces: microgels retain nearly spherical shapes on hydrophilic substrates but undergo pronounced spreading and flattening on hydrophobic surfaces, where favorable monomer-substrate interactions dominate.[54–56] Such substrate-induced shape changes redistribute polymer segments within the network and are therefore expected to modify the effective mechanical response of microgels, yet a direct, spatially resolved characterization of this effect has been lacking.

Here, we demonstrate how the substrate contact angle governs microgel shape, internal stiffness, and deposition patterns by employing atomic force microscopy (AFM) as powerful tool to probe microgels under hydrated conditions.[11,57–62] Specifically, force-volume measurements using sharp AFM tips are performed on microgels deposited on three differently silanized substrates, enabling the extraction of spatially resolved stiffness tomography maps and direct access to microgel internal structure and deformability. We find that microgels undergo pronounced deformation and stiffening on hydrophobic substrates, whereas they remain largely undeformed and soft on hydrophilic substrates. Moreover, microgels on hydrophilic surfaces show a strong sensitivity to their deposition history. Taken together, these results demonstrate that the substrate dictates microgel conformation and effective stiffness at interfaces, which in turn explains their drying behavior and resulting structures after deposition and drying.

## Results and Discussions

**Substrate wettability controls hydrated microgel shape and stiffness.** We investigate the mechanical properties of model poly(*N*-isopropylacrylamide) (PNIPAM) microgels[13,43] on three differently silanized substrates spanning hydrophobic to hydrophilic wettabilities: perfluorooctyltriethoxysilane (FOCTS) (Fig. 1A), n-octadecyltrimethoxysilane (ODS) (Fig. 1B) and 3-aminopropyltriethoxysilane (APTES) (Fig. 1C). All substrates exhibit similarly low surface roughness (<1 nm), excluding topographical effects and allowing a direct comparison of microgel behaviour as a function of substrate surface chemistry. The microgels adsorb *in situ* from aqueous suspension at solid-water interfaces to prevent any alterations that may occur during drying (Fig. 1D).

Force-volume AFM measurements allow reconstruction of the zero-force microgel topography, providing access to the unperturbed microgel shape under fully hydrated conditions. In addition, spatially resolved contact stiffness maps are obtained, enabling direct comparison of internal stiffness distributions within individual microgels.[48,63–65] Representative sharp tip force-volume AFM measurements of individual microgels adsorbed under fully hydrated conditions are presented in Fig. 1; additional measurements from repeated experiments are provided in the Supporting Information Fig. S1-6.

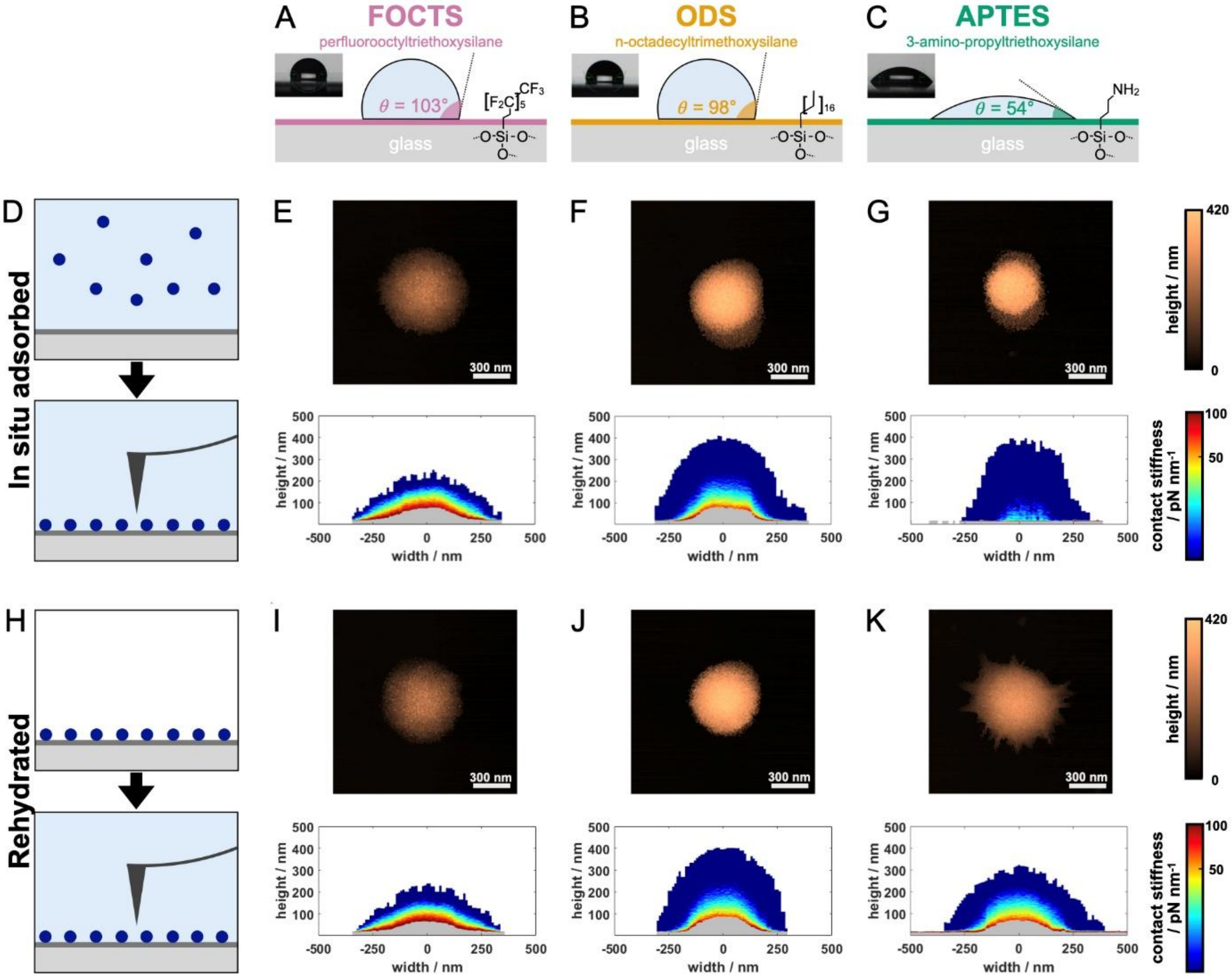

Figure 1: Corrected height images and stiffness tomography maps of the microgels on differently silanized substrates. Differently silanized substrates with indicated chemical structure and contact angle of water: FOCTS (A), ODS (B), and APTES (C). Microgels are analyzed at the solid-water interface after in situ adsorption (scheme: D, results: E-G), and after drying followed by rehydration (scheme: H, results: I-K). Corrected height images (top) and stiffness tomography profile through the middle of the microgels (bottom) for in situ adsorbed microgels on FOCTS (E), on ODS (F), and APTES (G) silanized substrates, and for rehydrated microgels on FOCTS (I), on ODS (J), and on APTES (K) silanized substrates. Color scale bar for corrected height images and stiffness tomography maps on the right. Grey color in the stiffness tomography maps reflect areas that are not accessible by the sharp AFM tip.

On the strongly hydrophobic FOCTS substrate, microgels adsorb in a highly flattened configuration with a large lateral footprint and pronounced stiffening toward the substrate (Fig. 1E). On ODS, the microgels show intermediate deformation and moderate near-substrate stiffening (Fig. 1F), whereas on hydrophilic APTES they remain comparatively tall and uniformly soft (Fig. 1G).

To quantify these substrate-dependent shape changes, we determined the lateral and vertical deformation ratios as a function of substrate contact angle (Figure 2). Lateral deformation is expressed as the contact diameter, $D_{contact}$, normalized by the hydrodynamic diameter, $D_h$ = 612 nm, whereas vertical deformation is expressed as the microgel height, $h$, normalized by $D_h$ (Figure 2, Table S1).

Microgels adsorbed in situ on FOCTS exhibit the largest lateral deformation ratio of approximately 1.2 and the smallest vertical deformation ratio of approximately 0.5, demonstrating pronounced spreading and flattening (Figure 2B,C, filled Symbols). Microgels on ODS show intermediate behaviour, with a lateral deformation ratio close to 1 and a vertical deformation ratio of approximately 0.7. Microgels on APTES exhibit lateral and vertical deformation <1, indicating a transition from a spherical to a cap-shaped morphology rather than lateral spreading.

Overall, increasing substrate hydrophobicity promotes lateral spreading, flattening, and internal stiffening under fully hydrated conditions. Thus, substrate wettability directly controls both microgel conformation and mechanical state. Interestingly, comparable near-substrate stiffening has been observed for electrostatically immobilized microgels, despite the fundamentally different nature of the microgel–substrate interaction.[66]

To further extend the wettability range, freshly plasma-cleaned substrates are used as a strongly hydrophilic reference, with water contact angles below 10°. However, these substrates cannot be included in the in-liquid force-volume AFM analysis because weak microgel-substrate adhesion leads to tip-induced displacement and pickup of the microgels, causing irreversible tip contamination. They are therefore used only as an additional strongly hydrophilic reference in the dry-state deposition experiments discussed below.

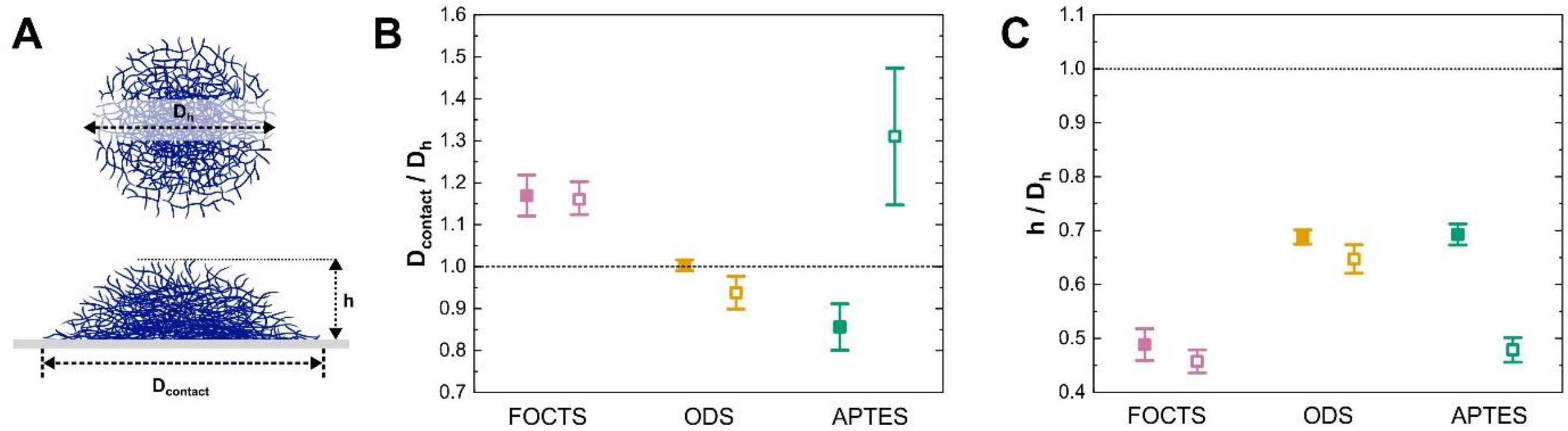

Figure 2. Lateral and vertical deformation ratios of microgels as a function of substrate contact angle. A) Schematic illustration of the hydrodynamic diameter, $D_h$, contact diameter, $D_{contact}$, and height, $h$. B) $D_{contact} / D_h$ and (C) $h / D_h$ for APTES (~55°), ODS (~100°), and FOCTS (~105°). Filled bars represent microgels adsorbed in situ, whereas open bars represent microgels after drying and rehydration.

**Drying and rehydration irreversibly flatten and stiffen microgels on hydrophilic substrates.** The response of adsorbed microgels to drying and subsequent rehydration is examined by repeating the AFM measurements (Figure 1H-K). On the hydrophobic FOCTS and ODS substrates, drying and rehydration do not induce detectable changes in microgel morphology or internal stiffness; the microgels retain their shape and remain mechanically stiff. This is also reflected in the deformation ratios, which remain nearly unchanged after rehydration (Fig. 2B,C, open symbols). In contrast, microgels adsorbed on the hydrophilic APTES substrate respond markedly differently: although they are initially soft and spherical cap-shaped under *in situ* conditions, drying followed by rehydration leads to pronounced flattening accompanied by a substantial increase in internal stiffness (Fig. 1K). The lateral deformation ratio $D_{contact} / D_h$ increases from approximately 0.85 to 1.3, while the vertical deformation ratio $h / D_h$ decreases from approximately 0.7 to 0.5 (Figure 2B,C, open symbols), demonstrating substantial spreading and flattening. The original microgel morphology and mechanical state are not recovered, indicating irreversible deformation, and filament-like protrusions extending toward the substrate are frequently observed after rehydration, consistent with partial detachment and restructuring during the drying process (Figure 1K, Figure S1-6).

We note that the lateral deformation ratios measured in liquid are considerably smaller than those reported for comparable single microgels in the dry state. This difference arises because dry-state AFM resolves both the densely cross-linked core and the loosely cross-linked corona, whereas in-liquid force-volume AFM primarily detects the denser inner region and remains less sensitive to the highly swollen corona.

Taken together, these observations show that substrate wettability governs not only microgel shape and stiffness at solid-water interfaces but also their sensitivity to deformation upon drying. Microgels adsorbed on hydrophilic substrates, which are initially soft, spherical cap-shaped, and weakly attached, undergo pronounced flattening and stiffening during drying and rehydration. In contrast, microgels on very hydrophobic substrates are already flattened and mechanically rigid under hydrated conditions and therefore remain largely insensitive to drying. These substrate-dependent mechanical states provide a basis for understanding how drying processes restructure microgel assemblies during deposition from interfaces.

**From single-microgel mechanics to drying-induced microgel assembly.** To relate the substrate-dependent microgel shape and internal stiffness revealed by force-volume AFM to their collective behavior after deposition and drying, we next examine microgel assemblies transferred from the air–water interface onto substrates by Langmuir–Blodgett deposition (Fig. 3A). Simultaneous compression and deposition were used to generate a gradient in microgel compression and interparticle spacing along a single substrate, which was subsequently analyzed by AFM and image analysis. We focus on surface pressures between 10 and 28 mN m$^{-1}$ (Fig. 3B), covering the initially hexagonally ordered regime, the previously reported isostructural transition region,[32,42–48] and the densely packed hexagonal regime at high compression. Importantly, all surface-pressure–area-per-particle isotherms collapse onto a common master curve, irrespective of the receiving substrate. Thus, any differences observed after transfer arise from substrate-dependent processes during or after deposition.

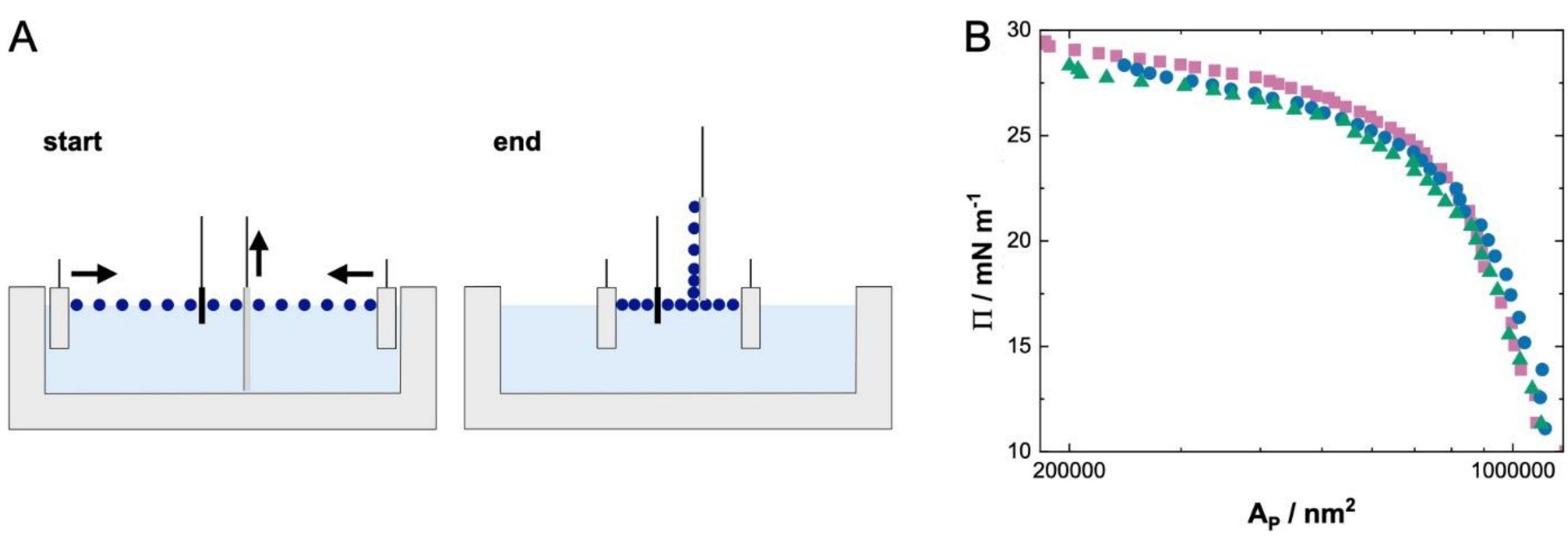

Figure 3: Langmuir-Blodgett trough and compression isotherms. Schematic illustration of a Langmuir-Blodgett trough experiment showing gradual transfer from the air-water interface (A), with the setup depicted at the start (left) and end (right) of deposition. Surface pressure, Π, versus measured area-per-particle, $A_P$, (B) – the latter determined from AFM image analysis after deposition. Data are shown for microgels deposited on plasma-cleaned (blue circles), APTES-functionalized (green–blue triangles), and FOCTS-functionalized (pink squares) substrates.

We first examine microgel monolayers transferred onto plasma-cleaned, highly hydrophilic substrates (Fig. 4A-D), which serve as a reference closely matching the conditions under which the isostructural transition was originally reported.[42,43] At low compression (11.1 mN m$^{-1}$), the microgels form a hexagonal non-close-packed lattice with corona-corona contacts. Between $\Pi$ = 25.5 and 27.4 mN m$^{-1}$, AFM micrographs reveal the previously reported isostructural transition, in which microgels with corona-corona contacts coexist with microgels in core-core contact and form clusters that grow upon further compression. Accordingly, the nearest-neighbor distance, *NND*, splits into two distinct values (Fig. 4E, blue and light-blue circles). At $\Pi \geq$ 28.6 mN m$^{-1}$, the monolayer remains a dense hexagonal lattice of core-core contacts, resulting again in a single *NND* (Fig. 4E).

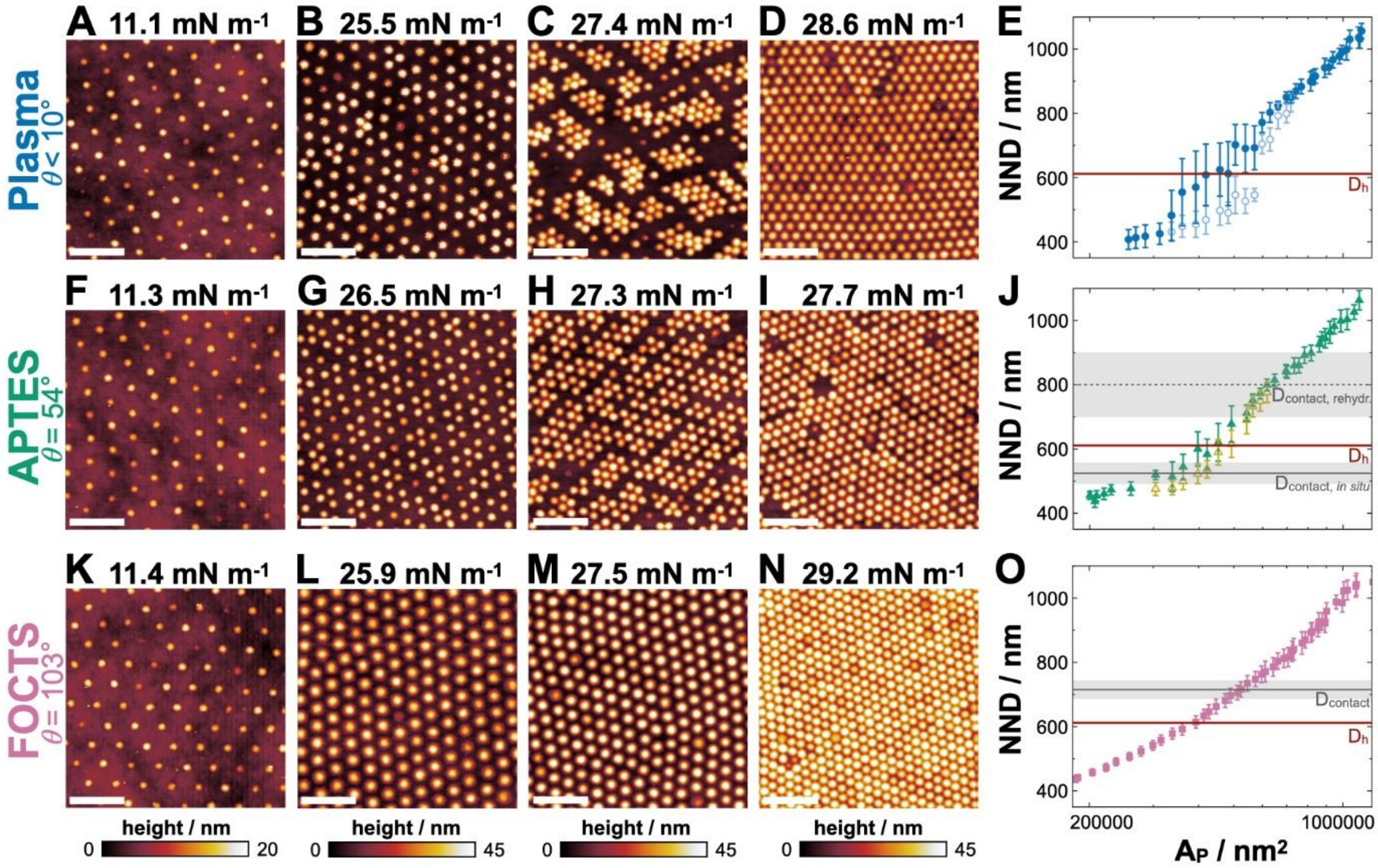

Figure 4: Drying-induced microgel assembly: AFM micrographs at different surface pressures of the microgel monolayers after deposition onto the three substrates (plasma-cleaned: A-D, APTES: F-I, and, FOCTS: K-N, scale bars: 2.5 μm and color bars below) and nearest-neighbor distance, *NND*, as a function of the area-per-particle, $A_P$, (plasma-cleaned: E, APTES: J, and, FOCTS: O). The *NND* splits-up into two values within the third compression regime for microgels deposited onto plasma cleaned (E, blue and light blue circles) and APTES functionalized substrates (J, green-blue and green-yellow triangles), while it continuously decreases – having only one value per $A_p$ – for microgels deposited on FOCTS silanized substrates (O, pink squares).

On the slightly less hydrophilic APTES substrate (Fig. 4F-I), a qualitatively similar but attenuated structural evolution is observed. At low compression (11.3 mN m$^{-1}$), the microgels form a hexagonal non-close-packed lattice with corona-corona contacts. In the transition regime ($\Pi$ = 26.5–27.3 mN m$^{-1}$), an isostructural transition remains detectable; however, both the splitting of the nearest-neighbor distance and the width of the transition regime are reduced compared with the more hydrophilic plasma-cleaned substrate (Fig. 4J, green-blue and green-yellow triangles).

On the strongly hydrophobic FOCTS substrate (Fig. 4K-N), microgels exhibit a distinctly different assembly behavior. Hexagonally ordered arrangements are observed at low and high compression (11.4 and 29.3 mN m$^{-1}$), but no isostructural transition occurs in the intermediate regime ($\Pi = 25.6 - 27.4$ mN m$^{-1}$). Instead, the nearest-neighbor distance decreases continuously with increasing compression, corresponding to lower area-per-particle values, $A_P$ (Fig. 4O, pink squares). Together, these observations demonstrate that, on the strongly hydrophobic FOCTS substrate, microgels undergo continuous packing without distinct corona-corona and core-core contact states, and that the isostructural transition is fully suppressed, in agreement with earlier observations for related microgel systems on hydrophobic substrates.[53]

Microgel deposition onto the hydrophobic ODS substrate could not be achieved by Langmuir-Blodgett transfer. Both withdrawal from and immersion through the air-water interface were tested, but neither approach resulted in detectable microgel transfer. This likely reflects the high water contact angle of ODS combined with the reduced spreading of the microgels on these substrates, which thereby prevents efficient wetting of the substrate and hinders transfer of the interfacial microgel monolayer.

**Mechanistic interpretation of the apparent isostructural transition.** Taken together, these results indicate that the substrate preconditions each microgel before drying, thereby determining how the transferred assembly reorganizes, as schematically summarized in Fig. 5. On hydrophilic substrates, microgels remain more spherical, soft, and only weakly immobilized at the substrate (Fig. 1G). Their larger effective height and lower stiffness make them more susceptible to immersion-capillary interactions during drying, while their weak immobilization allows lateral rearrangements into distinct corona-corona and core-core contact states (Fig. 4A-J). In contrast, hydrophobic substrates promote microgel spreading, flattening, and stiffening (Fig. 1E,F), thereby reducing the effective particle height, limiting further deformation, and immobilizing the microgels at the substrate. Consequently, capillary-driven deformation and lateral rearrangement during drying are suppressed, and the deposited particles retain the hexagonal arrangement (Fig. 4K–O).

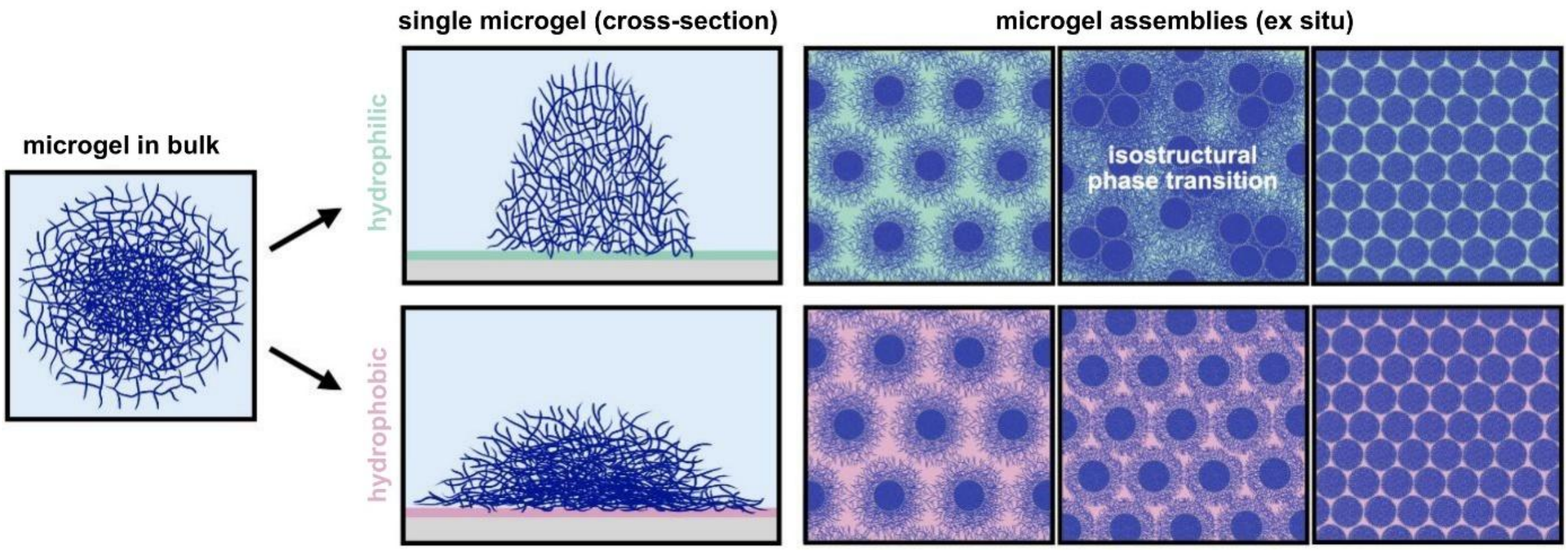

Figure 5: Schematic summary illustrating how the substrate contact angle controls single microgel's shape, internal stiffness, and the assembly behavior of microgel monolayers. Hydrophilic substrates preserve tall, soft, nearly spherical microgels and promote an isostructural phase transition, whereas hydrophobic substrates induce spreading, flattening, and stiffening of the microgels, leading to continuous packing without an apparent isostructural transition.

To assess whether this transition can be associated with a characteristic microgel length scale, we further compared the onset of the isostructural phase transition with the measured contact diameter, $D_{contact}$. For APTES, the splitting regime overlaps reasonably well with the measured contact diameters (Fig. 4J). However, corresponding $D_{contact}$ data are not available for the plasma-cleaned substrate. Moreover, on FOCTS, the NND reaches $D_{contact}$ without any apparent transition (Fig. 4O), indicating that this geometric coincidence alone cannot explain the transition. The present comparison therefore does not allow a general conclusion about the role of $D_{contact}$ in the isostructural transition. This question should be examined using a different model system in which $D_{contact}$ can be tuned systematically.

Our results corroborate previous reports that the apparent isostructural transition arises from transfer- and drying-induced rearrangements and further identify substrate-dependent microgel mechanics as a key factor controlling these rearrangements.[51,53] Nevertheless, the origin of the transition is not fully resolved. An isostructural transition has also been observed directly at oil–water interfaces in microgel-stabilized emulsions by cryo-SEM,[4,13,49,50] although the mechanism underlying its formation under these conditions remains unclear. It therefore remains possible that a comparable transition is already present at the air–water interface and is subsequently preserved, modified, or obscured during transfer and drying, depending on the strength of the microgel–substrate interactions. In this scenario, the apparently continuous packing pathway observed on hydrophobic substrates would reflect an altered ex situ signature of the interfacial transition rather than its complete absence. Direct structural measurements of the hydrated interfacial monolayer, for example by neutron or X-ray scattering during compression and transfer, will be required to test this hypothesis.

## Conclusion

In this work, we show that the contact angle of the substrate on which PNIPAM microgels adsorb directly controls their internal mechanical state. Depending on the substrate contact angle, the same microgels adopt markedly different shapes and stiffness profiles under fully hydrated conditions. In-liquid force-volume AFM reveals that adsorption on hydrophobic substrates induces pronounced spreading and flattening, accompanied by strong internal stiffening and immobilization on the surface. In contrast, adsorption on hydrophilic substrates preserves taller, more spherical, and softer microgels with smaller contact areas and weaker substrate attachment. Thus, substrate contact angle does not merely influence the drying process; it preconditions the microgels mechanically before drying begins.

Taken together, our results indicate that these substrate-dependent mechanical states determine how the monolayer responds during Langmuir-Blodgett transfer and drying. On hydrophilic substrates, the microgels remain tall, soft, and only weakly immobilized, making them susceptible to drying-induced immersion-capillary forces. As the water film recedes, these microgels can deform and rearrange laterally, producing distinct corona-corona and core-core contact states and thereby accounting for the apparent isostructural transition observed *ex situ*. On hydrophobic substrates, by contrast, the microgels are already flattened, stiffened, and immobilized under hydrated conditions. Their reduced height, limited deformability, and stronger surface immobilization suppress capillary-driven rearrangements, thereby preserving the hexagonal monolayer.

Overall, our results provide a mechanistic explanation for why the reported isostructural transition emerges under specific *ex situ* conditions. The substrate mechanically preconditions the microgels before drying begins by setting their shape, stiffness, contact area, and mobility. This preconditioning determines whether capillary forces during drying can reorganize the monolayer or whether the transferred arrangement is preserved. Substrate contact angle therefore provides a direct route to control the stability and restructuring of two-dimensional microgel assemblies during deposition and drying.

## Experimental Section

<u>Synthesis of microgels</u>

We use poly(*N*-isopropylacrylamide) (PNIPAM) microgels cross-linked with 2.5 mol% *N,N′*-methylenbisacrylamide (BIS) and a hydrodynamic diameter of 612 nm at 20 °C in water. The same batch of microgels has already been characterized in previous studies.[13,43]

The detailed description of the synthesis of the microgels is published elsewhere.[43] Briefly, the microgels were synthesized by surfactant-free precipitation polymerization in a 500 mL three-neck round-bottom flask equipped with a reflux condenser and stirrer. *N*-isopropylacrylamide (2.83 g, 25.0 mmol) and *N,N′*-methylenbisacrylamide (96.3 mg, 0.62 mmol) were dissolved in 249 mL of Milli-Q water, heated to 80 °C, and purged with nitrogen for 30 min. Polymerization was initiated by rapid addition of ammonium persulfate (14.3 mg, 0.063 mmol) dissolved in 1 mL of Milli-Q water and allowed to proceed for 5 h under nitrogen atmosphere. After cooling to room temperature, the microgels were purified by three cycles of centrifugation and redispersion, followed by dialysis against Milli-Q water for one week. Owing to the faster consumption of BIS compared to the monomer NIPAM during precipitation polymerization, the microgels possess a heterogeneous internal structure.[67]

<u>Preparation of silanized substrates</u>

Microscope glass coverslips (#2, 2.2 × 2.2 cm², Menzel Gläser) and silicon wafers (thickness 500 ± 50 nm, approx. 0.5 × 5.0 cm², MicroChemicals GmbH) were used as substrates. For silane modification, the substrates were first cleaned by two consecutive ultrasonication steps of 15 min each in technical-grade ethanol. They were then activated in a plasma oven (30 W, air flow 1.0 Nl $h^{-1}$; Zepto One, Diener Electronic GmbH & Co. KG) for 25 min. The silanes, (i) 1*H*,1*H*,2*H*,2*H*-perfluorooctyltriethoxysilane (FOCTS, >97%, TCI Deutschland GmbH), (ii) n-octadecyltrimethoxysilane (ODS, 90%, Thermo Scientific), and (iii) 3-aminopropyltriethoxysilane (APTES, 99%, Thermo Scientific), were dissolved either in hexane (≥97.0%, Sigma-Aldrich) or ethanol (≥99.8%, Fisher Chemical) at the following concentrations: FOCTS 2.5 vol% in hexane, ODS 5.0 vol% in ethanol, and APTES approx. 3.0 vol% in ethanol. The solutions were applied to the freshly plasma-activated substrates in a glass petri dish. After 30 min of incubation, the substrates were thoroughly rinsed with Milli-Q water (Purelab Flex, ELGA LabWater) and subsequently dried in an oven at 120 °C for 2 h.

Roughness of substrates via dry state AFM

Measurements were carried out on a Dimension Icon AFM equipped with a closed-loop system (Bruker Corporation; software: Nanoscope 9.4, Bruker Corporation). The silanized substrates were examined at the solid-air interface using OTESPA tips (MikroMasch, resonance frequency 300 kHz, nominal spring constant 26 N $m^{-1}$) operated in tapping mode. Surface roughness (RMS roughness) was quantified using the open-source software Gwyddion 2.67. Prior to analysis, all images were processed in Gwyddion 2.67 by aligning the scan rows and setting the zero level to the minimum z-value of each image.

AFM reveals that all substrates exhibit comparably low surface roughness, with root-mean-square values of 0.5 nm for FOCTS, 0.6 nm for ODS, and 0.4 nm for APTES.

Contact angle measurements

Static contact angle measurements were carried out using a DSA25 drop shape analyzer (Krüss GmbH). A 10 µL water droplet was placed on each silanized substrate at three different positions at room temperature. The contact angle was recorded over 10 seconds at a rate of one measurement per second. For FOCTS and ODS surfaces, the droplet contour was fitted using the Young-Laplace equation, while for APTES surfaces the circle-fit method was applied. For each substrate, the final contact angle value represents the average of all measurements across the three positions.

The highest contact angle is observed for the FOCTS substrate ($\Theta = 103\pm1°$), followed by the ODS substrate ($\Theta = 98\pm3°$), while the APTES substrate exhibits a significantly lower contact angle ($\Theta = 54\pm1°$). This systematic variation in surface wettability allows us to directly compare microgels behavior on hydrophobic versus hydrophilic substrates.

Sample preparation for AFM measurements in liquid

Microgels were investigated by atomic force microscopy (AFM) at the solid-water interface following two different deposition histories: (i) *in situ* adsorption and (ii) drying followed by rehydration.

**(i) *In situ* adsorption:** Microgels were adsorbed directly onto the differently silanized substrates under aqueous conditions. To achieve isolated microgels, the microgel stock solution (ca. 1.3 wt%) was diluted with Milli-Q water at substrate-specific ratios. For FOCTS substrates, 600 µL of a 1:20 (v/v) diluted stock solution was deposited, while for ODS and APTES substrates, 400 µL of a 1:3 (v/v) diluted stock solution was used. The substrates were placed in a customized AFM liquid cell prior to deposition, and the microgel solutions were incubated at room temperature for 30 min. After incubation, excess microgels were removed by rinsing with filtered (0.2 μm RC) Milli-Q water, and the liquid cell was immediately filled with water. The system was allowed to equilibrate for 45 min prior to AFM measurements at the solid-water interface. Throughout the experiment, the substrates remained fully wetted and were not exposed to drying before AFM measurements at the solid-water interface.

**(ii) Rehydration:** Following the *in situ* AFM measurements, the same substrates were removed from the liquid cell and dried in air for two weeks. To investigate the effects of drying and rehydration, the substrates were subsequently reintroduced into the customized liquid cell and rehydrated by filling the cell with filtered (0.2 μm RC) Milli-Q water at room temperature. The system was allowed to equilibrate for 45 min prior to AFM measurements at the solid-water interface.

Force-volume AFM measurements in liquid & automated force-distance curve analysis

AFM measurements were performed using a Nanowizard 3 atomic force microscope (JPK Instruments AG; software: Nanowizard Control Software v.6 JPK Instruments Bruker Corporation). The instrument was operated in quantitative imaging mode to acquire force-volume data. Microgels adsorbed on silanized substrates were investigated at the solid-water interface using MSNL-10 type E cantilevers (Bruker Corporation) with a nominal resonance frequency of 38 kHz, a nominal spring constant of 0.1 N $m^{-1}$, and a nominal tip radius of 2 nm.

Cantilever calibration was carried out using the thermal noise method. The deflection sensitivity was determined from the slope of force-distance curves recorded on the bare substrate and repeated several times to improve accuracy. Subsequently, the cantilever was retracted by 100 μm and the thermal noise spectrum was recorded using the thermal tune function of the AFM software. The spring constant was extracted from the corresponding power spectrum. To minimize uncertainties arising from calibration, all measurements were performed using the same cantilever.

Following calibration, overview images of larger surface areas were acquired. After identifying regions containing individually adsorbed microgels, force-volume measurements were performed over an area of 1.5 × 1.5 $\mu m^2$ with a resolution of 150 × 150 $pixels^2$. The force setpoint was set to 5.0 nN, the z-length to 1500 nm, and the pixel time to 50.0 ms. For each substrate type and deposition history, ten force-volume measurements were conducted to ensure statistical relevance.

Each force-volume dataset, comprising 22500 force-distance curves, was analyzed using custom-written Matlab scripts described in detail elsewhere.[48,63] Briefly, force-distance curves were first baseline-corrected, and the measured height was adjusted for cantilever deflection to obtain the true tip-sample distance. For each curve, a set of potential contact points was identified based on characteristic features of the data; the algorithm selected the most probable contact point and shifted the curve such that contact occurred at zero tip-sample distance. The height images were then corrected for sample indentation induced by the applied force setpoint by accounting for the difference in tip-sample distance between the contact point and the maximum applied force.

Stiffness tomography maps were subsequently extracted by calculating the local contact stiffness, defined as the slope of the force-distance curve in the contact regime, at different penetration depths. These data were assembled into a four-dimensional matrix containing the spatial coordinates (x, y, z) and the corresponding stiffness values. For visualization, planes intersecting the centers of individual microgels were constructed, with stiffness represented by color coding, while regions inaccessible due to the applied force limit of 5.0 nN are shown in grey.

Microgel dimensions and deformation ratios were determined from height profiles obtained by linear interpolation of the corrected AFM height data parallel to the fast-scan direction. The average profile of ten individual microgels was used to extract the maximum lateral and vertical dimensions, corresponding to the contact diameter, $D_{contact}$, and microgel height, $h$, respectively.

Deposition patterns: Langmuir-Blodgett deposition, AFM micrographs & automated image analysis

Compression isotherms and Langmuir-Blodgett (LB) deposition of microgel monolayers from the air-water interface were performed using a KSV NIMA LB trough made of PTFE and equipped with a platinum Wilhelmy plate (perimeter: 39.24 mm). The trough had an area of 243 $cm^2$ and a width of 7.5 cm. Prior to each experiment, all components of the LB trough were thoroughly cleaned by rinsing with ethanol and Milli-Q water, followed by drying with an air jet. The trough was then filled with Milli-Q water as the subphase.

Silanized silicon wafer substrates were mounted on a substrate holder and positioned at an angle of 90° or 45° relative to the interface. The substrates were immersed in the water subphase prior to deposition. A microgel suspension (0.1 wt%) mixed with 50 vol% ethanol as a spreading agent was gently deposited onto the air-water interface using a 100 µL pipette. After an equilibration time of 10 min, compression of the monolayer and deposition onto the substrate were initiated simultaneously. The barriers were moved at a constant speed of 1.1 mm $min^{-1}$, while the substrate was withdrawn at a dipper speed of 0.3 mm $min^{-1}$. After completion of the deposition, the substrates were allowed to dry for 5 min before removal from the trough. Owing to the fixed deposition angle and constant dipper speed, each position on the substrate could be directly correlated with the corresponding surface pressure. For ODS substrates, an additional inverse approach was tested, in which microgels were deposited during immersion into the water subphase. For each substrate, AFM images at the solid-air interface were acquired after 120 s of oxygen plasma etching at intervals of 1 mm using tapping mode with a size of $20 \times 20$ $\mu m^2$.

Image analysis was performed using a custom-written Python script based on the particle-tracking algorithm developed by Crocker and Grier.[68] Nearest-neighbors were identified by Voronoi tessellation. The resulting nearest-neighbor distance distributions were fitted with either a single Gaussian function or the sum of two Gaussian functions to extract the characteristic peak positions. The area-per-particle was determined by dividing the image area by the number of detected microgels.

## Acknowledgements

We thank the Münster Nanofabrication Facility (MNF) for providing access to the AFMs and especially acknowledge the support of the instrument contact persons, Riya Gupta and Steffen Lohrmann. We are also grateful to Björn Braunschweig granting us access to the plasma oven, the drop shape analyzer and the Langmuir-Blodgett setup. We further thank the mechanical and electrical workshops of the Institute of Physical Chemistry at the University of Münster for constructing the customized liquid AFM cell. We acknowledge M. Thomas Kuhnert for supporting scanning electron microscopy (SEM) measurements. We thank Keumkyung Kuk and Julian Ringling for valuable discussions. Financial support from the Deutsche Forschungsgemeinschaft (DFG) under grant number 573194956 is gratefully acknowledged.

## Data Availability

AFM raw data are available under DOI:10.17632/skzczkxwcx.1

## Author contributions

M.F.S.: conceptualization, validation, formal analysis, investigation, writing – original draft, writing – review & editing. S.S.M.: investigation, writing – review & editing. T.K.: software. S.S.: software. J.M.S.: software, formal analysis, investigation. S.K.K.: investigation. T.B.: methodology. M.K.: conceptualization, writing – review & editing. W.R.: conceptualization, writing – review & editing. M.R.: conceptualization, resources, writing – original draft, writing – review & editing, supervision, funding acquisition.